# THERMAL ANALYSIS OF CLIMATE REGIONS USING REMOTE SENSING AND GRID COMPUTING

Cristina Serban and Carmen Maftei

Faculty of Civil Engineering, Ovidius University of Constanta, Romania
serban.cristina@univ-ovidius.ro
cmaftei@univ-ovidius.ro

## ABSTRACT

*The analysis of climate regions is very important for designers and architects, because the increase in density and built up spaces and reduction in open spaces and green lands induce the increase of heat, especially in an urban area, deteriorating the environment and causing health problems. This study analyzes the Land Surface Temperature (LST) differences in the region of Dobrogea, Romania, and compares with the land use and land cover types using TM and ETM+ data of 1989 and 2000. As the analysis is performed on large data sets, we used Grid Computing to implement a service for using on Computational Grids with a Web-based client interface, which will be greatly useful and convenient for those who are studying the ground thermal environment and heat island effects by using Landsat TM/ETM+ bands, and have typical workstations, with no special computing and storing resources for computationally intensive satellite image processing and no license for a commercial image processing tool. Based on the satellite imagery, the paper also addresses a Supervised Classification algorithm and the computation of two indices of great value in water resources management, Normalized Difference Vegetation Index (NDVI), respectively Land Surface Emissivity (LSE).*

## 1. INTRODUCTION

In recent years, special attention was paid to global or regional temperature variability. Climate scientists have concluded that: (1) the earth's surface air temperature increased by about $0.6\,^{0}$C during the 20th century, and (2) the increase of temperature was highest in 1990s [11]. Heat emissions into the atmosphere are a great contributor to global warming and climate change. Among sources of heat producing are many human activities, from industrial processes to household activities and all forms of transportation.

One of the key sources of heat is the Urban Heat Island (UHI) phenomenon which traps heat in thermal mass like concrete, asphalt, bricks, stones and black roads which absorb, store and then re-emit this heat to the urban air at night. Thus, the temperature in a city is, on average, 5 to 8 $^{o}$C higher than the surrounding non-urban areas, the difference generally being greater at night than during the day.

The various land cover types such as high, medium or low density built up spaces, sparse or dense vegetation, barren land or water bodies, the land use types such as the commercial, industrial, institutional, residential and open spaces, the number of inhabitants and the size of a city, the thermal properties of surface materials, the lack of evapotranspiration in urban areas and some meteorological factors like wind speed, air temperature, humidity and amount of precipitation, further enhances the heat. Thus it is essential to study the thermal environment of a region to optimize its development plans and to better design the green lands.

The LST (Land Surface Temperature) has a direct impact on air temperature and it is also one of the key parameters in the physics of land-surface processes [29], as another index, NDVI (Normalized Difference Vegetation Index).

LST and NDVI are required for a wide variety of scientific studies, from climatology to hydrology, to ecology and bio-geology. For example, they can be used in modelling large scale hydrological system, the greenhouse effect, the effects upon glaciers, ice sheets, permafrost and the vegetation in Earth's ecosystems, in agricultural applications (such as drought monitoring, determining frost damage in fruit crops, evaluating water requirements for crops during summer when they are prone to heat stress, or computing a water deficit index [19]), in forecasting the soil freezing or in analyzing heat islands in urban areas. Many other indices can be calculated based on LST and NDVI. For example, the slope of the LST/NDVI curve has been related to the soil moisture conditions [28].

Due to satellite images large size – up to 1 GB, in order to remotely estimate LST and NDVI using satellite imagery it is desirable to distribute the processing of satellite images over a heterogeneous network of computers, where each of them contributes to a faster result according to its capabilities. In this context, Grid computing may be a solution.

Grid Computing is an emerging technology that provides access to computing power and data storage capacity distributed over the globe. Grid computing (or the use of Computational Grids) is the use of multiple computers to solve a single problem at the same time – usually a scientific problem that requires a great number of computer processing cycles or access to large amounts of data.

Grid computing not only provides the resources that allow the scientists and researchers to manage vast collections of data, it also allows this data to be distributed all over the world, which means scientific teams can work on international projects from the comfort of their own laboratories, sharing data, data storage space, computing power, and results. Together, researchers can approach bigger questions than ever before: from disease cures and disaster management to global warming.

Grid technology combines high performance capability and high throughput computing, data intensive and on-demand computing, and collaborative computing through a set of service interfaces based on common protocols. Grid computing can be thought of as distributed and large-scale cluster computing and as a form of network-distributed parallel processing. It can be restricted to the network of computer workstations within an organization or it can be a public collaboration, in which case it is also sometimes known as a form of peer-to-peer computing.

Grid computing is driven by four major principles: resource sharing and efficient use, secure access, and interoperability between different grids, achieved by the adoption of open standards for Grid development. Many studies have been conducted to develop efficient and secure grid environments, with specific algorithms to be run on [16], [17].

A Grid computing system generally comprises of two types of grid systems namely, Computational Grids and Data Grids. The key role of the Computational Grids is to provide solutions to the complex scientific or engineering problems that use complex, computational intensive problem-solving algorithms, for example, weather forecasting, medical diagnoses, satellite image processing etc. These applications use services of a Data Grid in order to access the distributed data sets in a distributed networked environment. So, the key role of a Data Grid is to provide the backbone on which a Computational Grid performs its operations.

In this study, we describe a service for using on Computational Grids, that highlights the variations in the thermal environment that exist throughout the Dobrogea region due to different land cover types. For this purpose, we used Landsat TM and ETM+ data of 1989 and 2000 to firstly retrieve the Land Surface Temperature (LST) from bands 6 and 6.1, respectively. Secondly, we perform a Supervised Classification with the parallelepiped algorithm, in order to map land use/cover patterns. Comparing the LST images and the land cover classification images, one can identify the specific locations of heat islands within a region.

This paper is organized in 6 sections. The first section is Introduction and the second presents the Related Work. In section 3 are detailed the Input Data Sets and the Methodology. Next, in section 4, we describe the service that uses the Computational Grid and the experimental results. Conclusion and further work are approached in section 5.

## 2. RELATED WORK

The studies of thermal environment of a geographic area are traditionally performed with in situ measurements of air temperatures and ground meteorological data [4], [25]. Ground observations do not provide end-users with required spatial and temporal data, representative for large areas; continuous monitoring of specific environments is hampered by the sparse and/or irregular distribution of meteorological stations, the difficulties in performing ground surveys and the complexity of interpolating existing station data. Remotely sensed surface temperature and land use/cover types over an entire large area is therefore of major interest for a variety of environmental and ecological applications [4], [5], [10], [15], [18], [20], [21], [26], [28].

Typically, the thermal environment of an area is performed using techniques of remote sensing combined with a Geographic Information System (GIS) and satellite imagery derived from different spatial earth observation programs such as Landsat, MODIS, EUMETSAT, IRS, Ikonos, QuickBird, OrbView, etc. Therefore, we consider that the service described in this paper will be greatly useful and convenient for those who are studying the ground thermal environment and urban heat island effects by using Landsat TM/ETM+ images, and have typical workstations, with no special computing and storing resources for computationally intensive satellite image processing and no license for a commercial image processing tool.

## 3. CASE STUDY

### 3.1. Study Area

The Dobrogea region was selected as the case study area due to its high exposure to aridity, drought and even desertification phenomenon, all of these being increased by changes in land cover types occurred over time. Dobrogea is a region situated in the South – East of Romania, between the Black Sea and the lower Danube River – Fig.1.

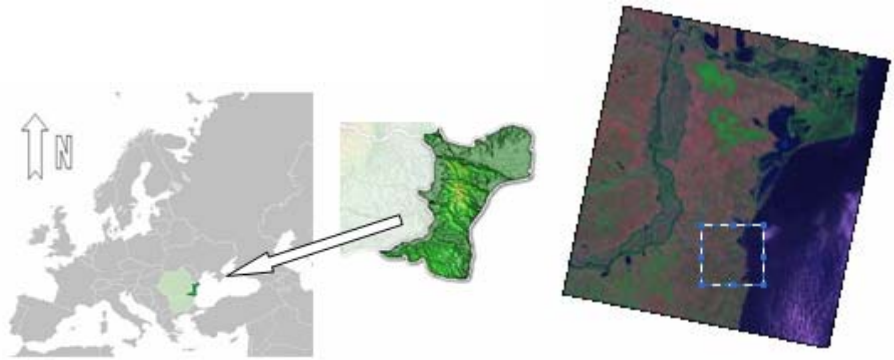

Figure 1 Dobrogea region

## 3.2. Input Data Sets

Input Data consists of two sets, satellite data set, respectively, ground meteorological data set.

### 3.2.1 Satellite Data

In this study, we used two subsets of Landsat TM and ETM+ images dated 20th August 1989 and 7th June 2000 – Fig.2, that cover the same area to facilitate comparison of images. The images are in geo-tiff format and were downloaded from [14]. Both images had good weather conditions without or little clouds in the study area. A radiometric calibrations (atmospheric corrections) was the pre-processing step that was taken. The bands TM 6, ETM+ 6.1, TM/ETM+ 3 and TM/ETM+ 4 were analyzed with respect to LST, whereas the other bands (TM 4, 5, 1, and ETM+ 7, 4, 2) were used for Supervised Classification.

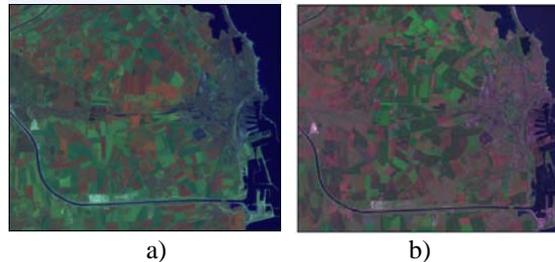

a)                                    b)

Figure 2 Landsat ETM+ "false color" image(bands combination: 742):a) year 1989 b) year 2000

### 3.2.2 Ground meteorological data

In addition to satellite data, this study needed a ground meteorological parameter (the total atmospheric water vapour content - w), used in LST estimation algorithm, which was obtained from the weather stations located on the study area. The weather conditions prevailing on 20[th] August 1989 and 7[th] June 2000 are shown in Table 4.

## 3.3. Methodology

### 3.3.1 Satellite Data Pre-processing

Satellite data pre-processing comprise of radiometric calibrations (atmospheric corrections) for TM/ETM+ bands 3 and 4. These bands are used to retrieve NDVI (Normalized Difference Vegetation Index) values on which the LST estimation algorithm is based on. It is possible to obtain NDVI values from at-sensor or TOA (Top of Atmospheric) reflectivities, called as NDVITOA, but it is more accurate to atmospherically correct the TOA values in order to obtain at-surface reflectivities and, in this way, estimate NDVI values more representative of the natural surfaces, called as NDVIsurf.

In this study we applied an atmospheric correction based on image data, developed by [22], its main advantage being that the data necessary in order to carry out the atmospheric correction are obtained from the image itself. The at-surface reflectivity is calculated with the following equation:

$$\rho_{surf} = \frac{\pi(L_{sensor} - L_p)d^2}{E_0 \cos\theta_z T_z}$$

(1)

where $L_{sensor}$ is at-sensor radiance, $T_z$ is the atmospheric transmissivity between the sun and the surface ($T_z \approx \cos\theta_z$ [23]), θz is the zenithal solar angle, E0 is the spectral solar irradiance on the top of the atmosphere [7], d is the Earth–Sun distance [7], and $L_p$ is the radiance resulted from the interaction of the electromagnetic radiance with the atmospheric components (molecules and aerosols) that can be obtained according to:

$$L_p = L_{min} - L_{1\%}$$

(2)

where $L_{min}$ is the radiance that corresponds to a digital count value for which the sum of all the pixels with digital counts lower or equal to this value is equal to the 0.01% of all the pixels from the image considered. $L_{min}$ was calculated through DOS (Dark Object Subtraction) technique [23], while the term L1% is given by

$$L_{1\%} = \frac{0.01\cos\theta_z T_z E_0}{\pi d^2}$$

(3)

### 3.3.2. LST Estimation

Currently, there are three different methods to retrieve LST directly from TM/ETM+ thermal band, all described in [9]. The method implemented in this study, the Jimenez-Munoz and Sobrino's algorithm, is the one which, firstly, needs the fewest ground weather data, (only one parameter), unlike the others which need in situ atmospheric profile launched simultaneously with the satellite passes, respectively, two ground weather parameters. Secondly, the Jimenez-Munoz and Sobrino's algorithm seems to give better results in given situations [8].

In order to estimate LST, the following steps were taken:

- Calculate NDVI, by applying the formula

$$NDVI = \frac{band4 - band3}{band4 + band3}$$

(4)

- Calculate Land Surface Emissivity (LSE), based on NDVI values. It makes use of the NDVI Thresholds Method—NDVITHM [8], by applying the formula

$$LSE = 1.0094 + 0.047 * \ln(NDVI)$$

(5)

when the NDVI value ranges from 0.157 to 0.727. When the NDVI value is out of the range (0.157–0.727), the corresponding input LSE constant values are used [8].

- Calculate LST, by applying the Jimenez-Munoz and Sobrino equation:

$$T_s = \gamma[\varepsilon^{-1}(\psi_1 L_{sensor} + \psi_2) + \psi_3] + \delta$$

(6)

with

$$\gamma = \{\frac{c^2 L_{sensor}}{T_{sensor}^2}[\frac{\lambda^4}{c_1}L_{sensor} + \lambda^{-1}]\}^{-1}$$

(7)

and

$$\delta = -\gamma L_{sensor} + T_{sensor},$$  (8)

where:

- $\varepsilon$ - LSE,

- $L_{sensor}$ - at-sensor radiance,

  $$L_{sensor} = gain * DN + bias$$  (9)

  where

  gain - band-specific rescaling gain factor

  bias - band-specific rescaling bias factor

  $DN$ - the digital number of a pixel,

- $T_{sensor}$ - brightness temperature,

  $$T_{sensor} = \frac{K_2}{\ln(K_1 / L_{sensor} + 1)}$$  (10)

- K1, K2 – calibration constants [7]

- $\lambda$ - the effective wavelength,

- $c_1 = 1.19104 * 10^8 W \mu m^4 m^{-2} sr^{-1} \; c^2 = 14387.7 \mu m K$,

- $\psi_i, i = \overline{1..3}$ - atmospheric parameters, obtained as functions of the total atmospheric water vapour content (w) according to the following equations particularized for TM/ETM+ 6 data:

  $$\psi_1 = 0.1471w^2 - 0.15583w + 1.1234$$
  $$\psi_2 = -1.1836w^2 - 0.37607w - 0.52894$$  (11)
  $$\psi_3 = -0.04554w^2 + 1.8719w - 0.39071$$

### 3.3.3 Supervised Classification

The supervised image classification method used is the parallelepiped algorithm or the "box decision rule classifier". The parallelepiped classifier uses intervals of pixels' values to determine whether a pixel belongs to a class or not. Classes are given by an image analyst, and represent an area of known identity delimited on the digital image, usually by specifying the corner points of a rectangular or polygonal area using the line and column numbers within the coordinate system of the digital image. More about the parallelepiped algorithm is to be found in [6].

In this study, seven land cover classes were classified in both the TM (1989) and ETM+ (2000) images by using bands TM 4, 5, 1 and ETM+ 7, 4, 2. The classes include: high density built up spaces, medium density built up spaces, low density built up spaces, dense vegetation, sparse vegetation, water and the barren land.

## 4. Approach

Our service meets the requirements of a virtual organization (VO) member who has access to a local database of large satellite images and wants to apply several satellite image processing operations in order to analyze the thermal environment of a region. The operations to be performed are implemented in special client's codes and are to be run on the Computational Grid.

Due to the large size of a satellite image (up to 1 GB), the full image transfer should be avoided. Therefore, a satellite image will be split into a number of sub-images equal with the number of workstations of the Grid Cluster. The image processing algorithms will also be split into independent tasks that can be performed in parallel and that are requiring similar computing effort.

The design applied is called the Split and Aggregate design which allows to parallelize the process of task execution gaining performance and scalability.

Fig. 3 (http://www.gridgain.com/) shows the logical steps on a Computational Grid: a Grid task splits into Grid jobs that are executed on Grid nodes, the results of the jobs are then aggregated into one, namely the Grid task result.

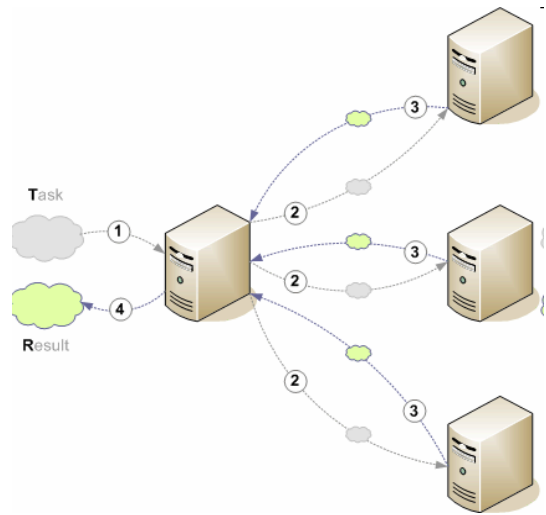

Figure 3 Split and Aggregate design: 1. Grid task execution request; 2. Grid task splits into Grid jobs; 3. Result of job execution; 4. Aggregation of job results into Grid task result

The following components are needed:

Table 1 Files used in the study

| Satellite image files | Filename | Type | File Size |
|---|---|---|---|
| TM 1989 | Bands: 3, 4 6 | tiff | 3000x3000 |
|  | "False color" image: 451 | tiff | 3000x3000 |
| ETM+ 2000 | Bands: 3, 4 61 | tiff | 3000x3000 |
|  | "False color" image: 742 | tiff | 3000x3000 |

- at the user's node: the satellite images (Table 1), the client's codes and some minimal facilities to access Grid infrastructure,

- at remote computing nodes: the Grid middleware which allows the execution of client's codes.

The client's code consists of three components – Fig.4:

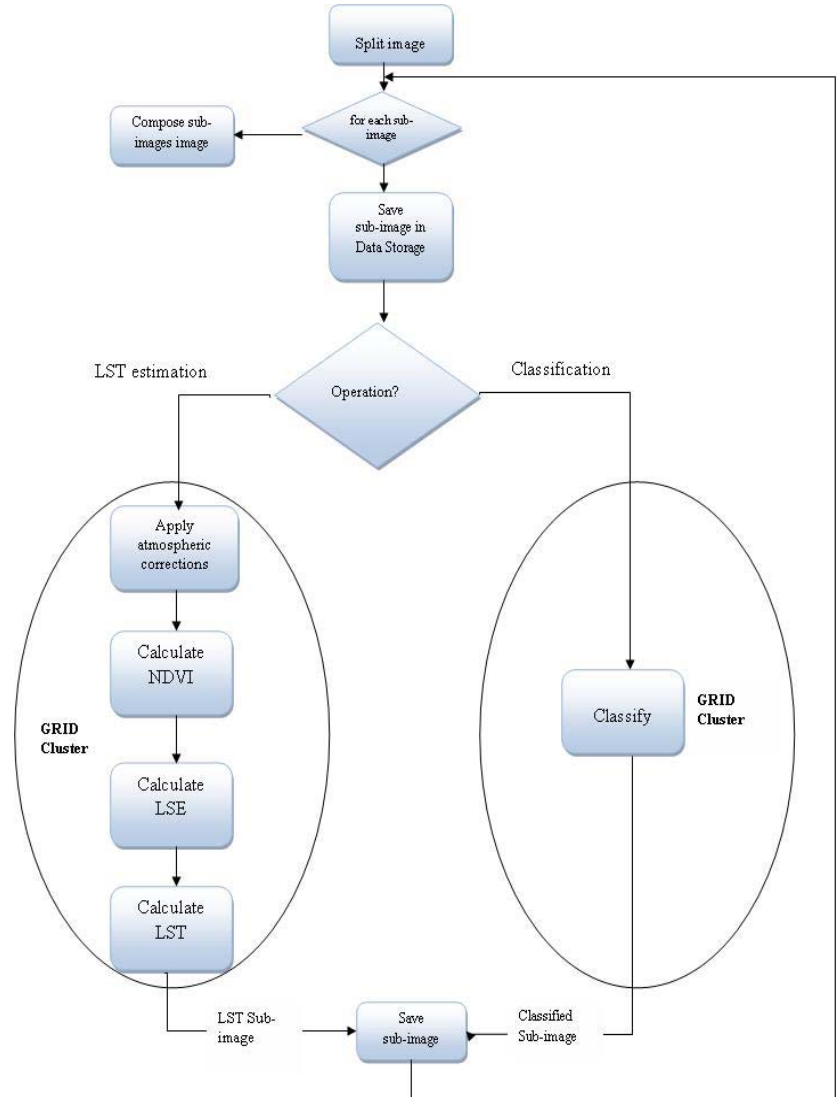

Figure 4 The logical processing steps for each sub-image

- the Splitter that takes a satellite image and split it into a number of sub-images;

- the ImageProcessor, which in turn consists of two components:

    o the ImageClassifier that receives a sub-image, applies the classification algorithm described in Section 3.3.3 and produces the output classified sub-image

    o the ImageLSTEstimator that receives a sub-image, applies the processing algorithms described in Sections 3.3.1 and 3.3.2 and produces the output LST sub-image

- the Composer that merges the resulting sub-images.

The Splitter and the Composer programs run only at the code's site where the large satellite images are residing. The ImageProcessor and the sub-images are submitted for processing on the Computational Grid.

The user uploads the image files and submits the jobs to the Grid. After the successful finish of the jobs, the user can download the resulting images.

The service works as follows – Fig.5:

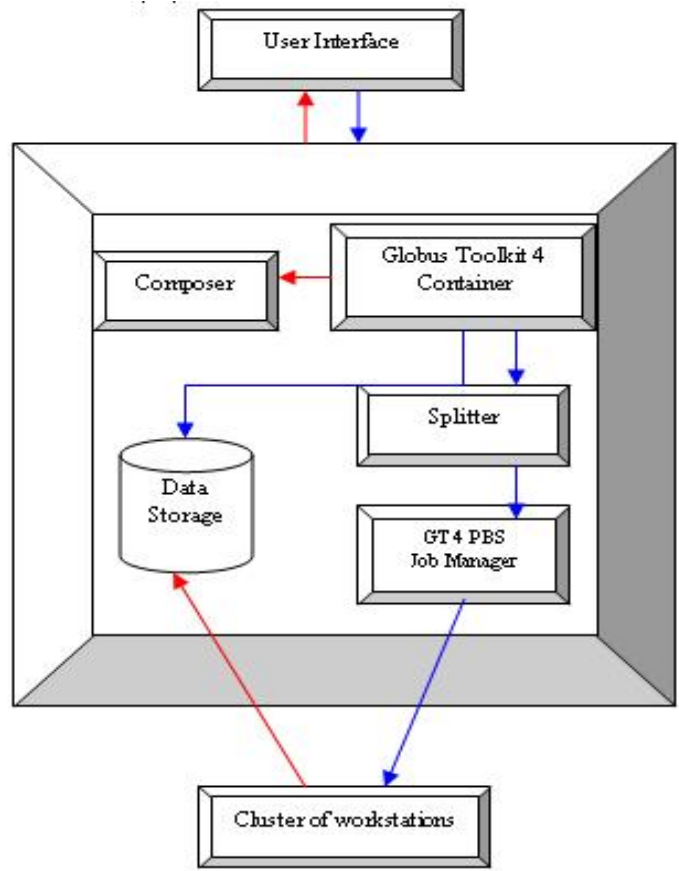

Figure 5  The components of the service and their interactions

- the user uploads the files using GridFTP and chooses an ImageProcessor operation (LST estimation or classification);

- the file(s) are transferred to code site;

- the Splitter code is called and the smaller pieces of image(s) are produced as well as the files needed by PBS to launch the ImageProcessor operation on each sub-image;

- The Job Manager of Globus Toolkit 4 take over the files and interpret them and finally PBS sends the jobs on the cluster of workstations;

- After the job executions the output file is stored on the code site;

- The user can access the output file through the user interface.

The client's code were written in Java and tested using Python scripts – Fig.6 and Fig.7 and on the Computational Grid provided by Globus Toolkit 4. Also, a Web-based client interface – Fig.8, for a service that launches the codes has been built using JSP, Tomcat/5.5 and MySQL.

```
submitCommandTemplate = "${GLOBUS_LOCATION}/bin/globusrun-ws -submit -s -batch -o subimage%02d.epr -F %s -Ft %s -f

# choose randomly the PBS or SGE factory
factoryType = random.choice(['PBS','Fork'])
factory = factoryCatalog[factoryType]

jobDescriptionFilePath = createGridJobDescriptionFile(subimageNumber)

submitCommand = submitCommandTemplate % (subimageNumber,factory,factoryType,jobDescriptionFilePath)
print "Preparing to submit job to grid..."
print "Submit command: %s" % submitCommand
try:
        job = popen2.Popen3(submitCommand, capturestderr = True)
        jobOut = []
        jobErr = []
        ret = job.poll()
        while ret == -1:
                jobOut.extend(job.fromchild.readlines())
                jobErr.extend(job.childerr.readlines())
                time.sleep(1)
                ret = job.poll()
except Exception, e:
                msg = "Error while submitting GRAM WS job: %s" % e
                raise RuntimeError, msg
if ret != 0:
        errorString = "".join(jobErr)
        msg = "Error while submitting grid job: %s" % errorString
        raise RuntimeError, msg
encodingGridJobCompletion[subimageNumber] = False
print "Grid job submitted to factory type %s" % factoryType
print

# This is a template for the job description file that will be
# used to submit each grid job.
jobDescriptionTemplate = """\
<job>
    <executable>/usr/local/sun/jdk/bin/java</executable>
```

Figure 6 Python script for code testing

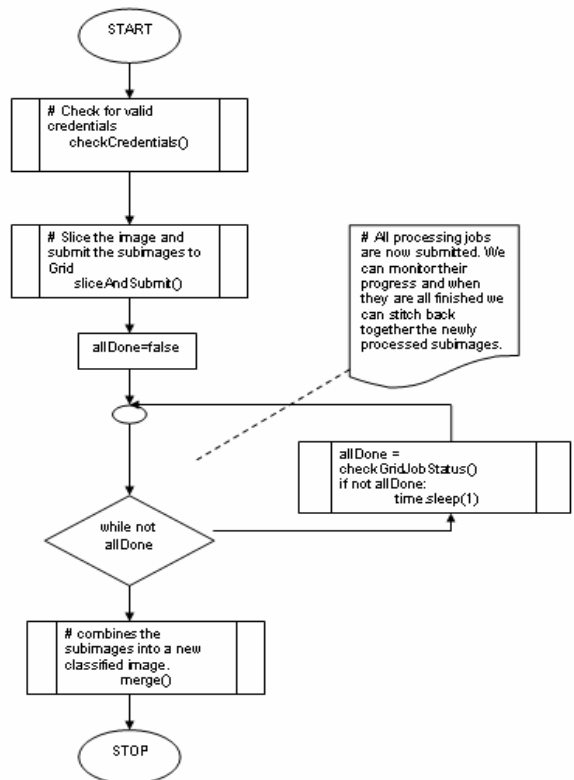

Figure 7 Python script flowchart

Figure 8 The user interface

The testing environment contains 4 PC nodes (Intel P4, 2.4 GHz, 1GB DDRAM) connected at 100 Mbps and allows processing images of size up to 10 MB. The tests that we performed proved that the presented application is efficient in terms of computation time and easy to use – Table 2.

Table 2 Response time for the sub-images processing

| No. of nodes | Response time (s) | File Size (Mb) |
|---|---|---|
| 1 | 8.5603 | 6.4 |
| 4 | 3.7621 | 1.6 |

The application output data consist of the files described in Table 3.

Table 3 The output data

| File | Description |
|---|---|
| .txt | Text files with LST values ($^0$C)[*] |
| .tiff | LST Map[*] |
| .tiff | Classified Image[*] |

[*] one file for each of the year studied

Based on the Supervised Classification of the 1989 image and the 2000 image (Fig. 9 and Fig. 10), the areas of dense built up spaces and barren land have increased, while the area of green space has been reduced to a greater extent.

The LST maps (Fig.11 and Fig.12) show that the LST values for the study area vary between $22^0$C la $48^0$C, which is consistent with the ground meteorological data measured by Constanta station (measurements are taken every 6 hours) (Table 4). The highest LST values are obtained for the building area ($40^0$C - $48^0$C). For the area covered by vegetation, the LST varies between $31^0$C and $37^0$C, the higher values being obtained for the sparse vegetation spots. The lowest LST values were observed for water bodies.

Comparing the LST maps and the land cover classification images, the relationship between the land cover classification and the land surface temperature can be clearly understood. The hot spots are mainly concentrated in the urban areas and the barren land.

For the supervised classification procedure, data was collected using field observation and local knowledge of the area. Based on the knowledge of the study area features, 7 training classes were defined. The parallelepiped classifier presents good performance, demonstrated by the results of the confusion matrix and an overall accuracy of 89%. Depending on how the training data is defined, the accuracy of the classifier can be improved up to 95-96%.

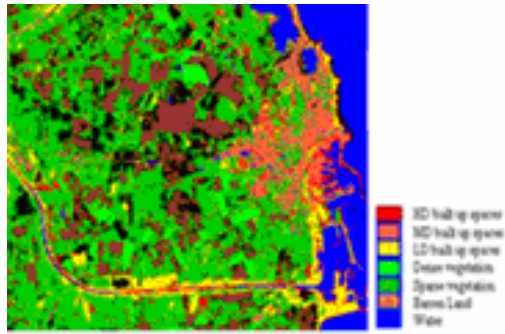

Figure 9 Classified Image Dobrogea region and Constanta city, 1989

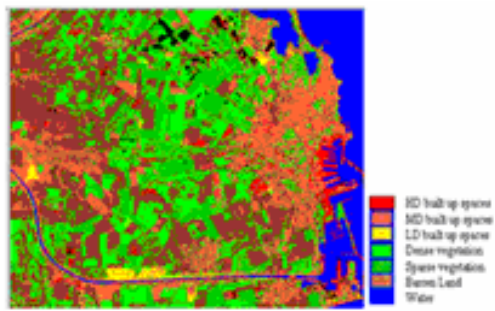

Figure 10 Classified Image Dobrogea region and Constanta city, 2000

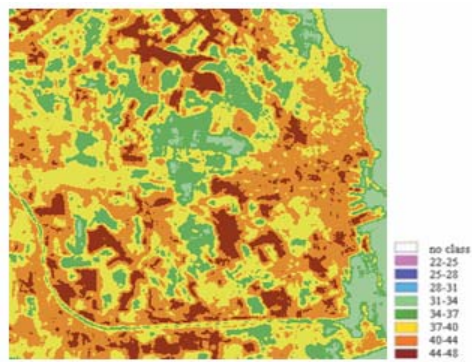

Figure 11 LST Image Dobrogea region and Constanta city, 1989

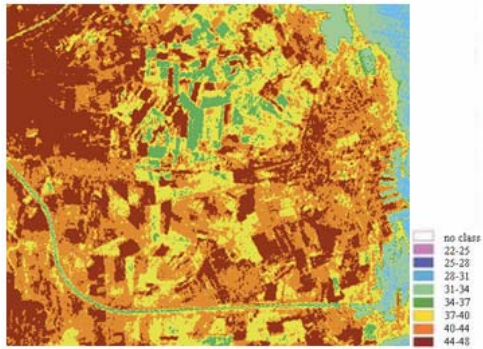

Figure 12 LST Image Dobrogea region and Constanta city, 2000

According to the meteorological data acquired from the Constanta weather station and the data computed (Table 4, Fig.13 and Fig.14), the temperature error between the actual mean ground surface temperature at the time when satellite passed and the remote sensed mean surface temperature is considered an acceptable one.

Table 4 Comparison between actual mean LST and remote sensed mean LST

| **Date:** | **Aug 20, 1989** | **Date:** | **June 7, 2000** |
|---|---|---|---|
| **Satellite Overpass Time** | | **Satellite Overpass Time** | |
| GMT | 7.36 | GMT | 8.32 |
| Local Time | 10.36 | Local Time | 11.32 |
| **Weather Station Data** | | **Weather Station Data** | |
| Air Temp ($^o$C) | 26 | Air Temp ($^o$C) | 23.3 |
| Humidity (%) | 62 | Humidity (%) | 82 |
| w (g/cm$^2$) | 2 | w (g/cm$^2$) | 2.3 |
| Mean  LST ($^o$C) | | Mean  LST ($^o$C) | |
| At 7 am: | 25.8 | At 7 am: | 21.2 |
| At 13 am: | 49.8 | At 13 am: | 48 |
| At 10.30 am: | 39.8 | At 11.30 am: | 42.31 |
| **Remote sensed LST** | | **Remote sensed LST** | |
| Mean LST ($^o$C) | 38.64 | Mean LST ($^o$C) | 41.58 |

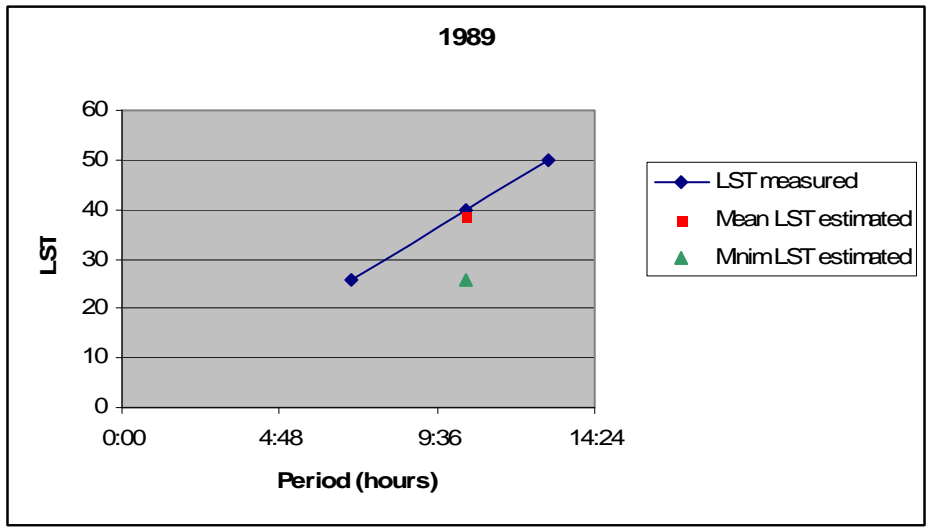

Figure 13 The comparison between measured and estimated LST values for year 1989

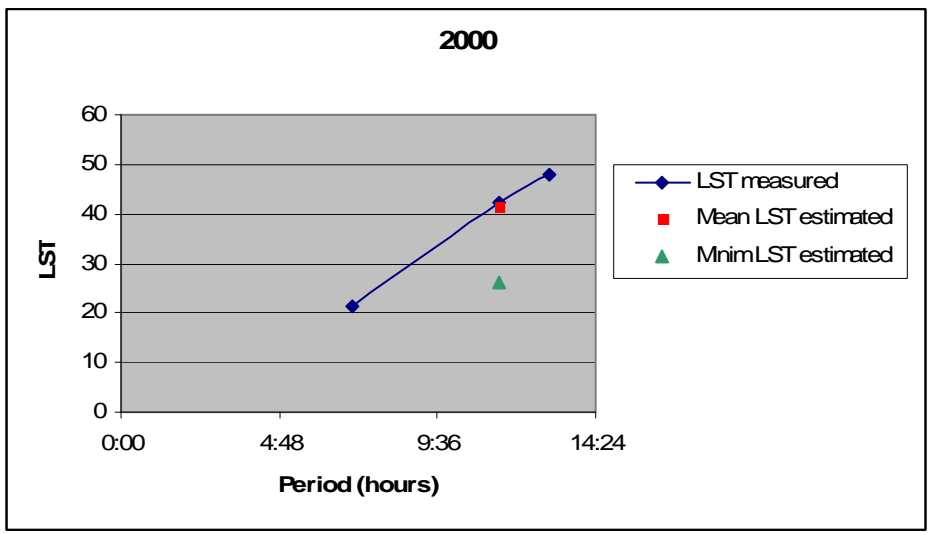

Figure 14 The comparison between measured and estimated LST values for year 2000

## 5. CONCLUSIONS

Remote sensing-based studies can aid designers and architects in providing better plans for building design and landscaping of area developments that are useful in minimizing the heat accumulation and retention by surfaces. Grid computing will assist those who want to do such studies but don't have the computational resources they need.

Also, remote sensing can assist in improving the estimation of various parameters of great value in water resources management in large cultivated areas and agricultural drought monitoring, Normalized Difference Vegetation Index (NDVI) and Land Surface Temperature (LST) are just two of them. The results of this study regarding these parameters are promising and show that the approach can derive reasonable estimates for NDVI and LST. Our future work will focus in developing an irrigation decision support system created with an internet server-based architecture to collect and present information from remote sources at one location. It will be tested in Dobrogea, and will work in cooperation with regional weather stations. Spatial drought monitoring and assessment over the Dobrogea region are also taken into consideration.

## ACKNOWLEDGEMENTS

The work described in this paper was funded by Grant PN II IDEI ID 262

## REFERENCES

[1]     B. Jacob, M. Brown, K. Fukui, N. Trivedi, "Introduction to Grid Computing", SG24-6778-00. http://ibm.com/redbooks, 2005

[2]     D. Petcu, Arhitecturi si tehnologii Grid, Editura Eubeea, 2006

[3]     D.R. Streutker, "A remote sensing study of the urban heat island of Houston, Texas", International Journal of Remote Sensing, 23, 2595 – 2608, 2002

[4]     D.R. Streutker, "Satellite measured growth of the urban heat island of Houston, Texas", Remote Sensing of Environment, 85, 282 – 289, 2003

[5]     E. Boegh, H. Soegaard, N. Hannan, P. Kabat, L. Lesch, "A remote sensing study of the NDVI – Ts relationship and the transpiration from sparse vegetation in the Sahel based on high resolution satellite data", Remote Sensing of Environment, 69, 224 – 240, 1998

[6]     ENVI Tutorial, www.ittvis.com/Envi/docs/tutorials/Classification Methods.pdf

[7]     G. Chander, B L. Markham, D.L. Helder, "Summary of current radiometric calibration coefficients for Landsat MSS, TM, ETM+, and EO-1 ALI sensors", Remote Sensing of Environment 113 (2009) 893–903, 2009

[8]     J. Sobrino, N. Raissouni, Z. Li, "A comparative study of land surface emissivity retrieval from NOAA data", Remote Sensing of Environment, 75, pp. 256-266, 2001

[9]     J. Sobrino, J.C. Jime´nez-un˜oz, L. Paolini, "Land surface temperature retrieval from LANDSAT TM 5", Remote Sensing of Environment 90 434–440, 2004

[10]    J. Voogt , T.R. Oke, "Thermal remote sensing of urban climates" Remote Sensing of Environment 86, 370–384, 2003

[11]    J.T. Houghton, Y. Ding, D.J. Griggs, M. Noguer, P.J. van der Linden, X. Dai, K. Maskell, C.A. Johnson (Eds.), "Climate Change, The Scientific Basis". Cambridge University Press, Cambridge, UK., 944 pp, 2001

[12]    J. Wu, D. Wang, M.E. Bauer, "Image-based atmospheric correction of QuickBird imagery of Minnesota cropland", Remote Sensing of Environment 99 (2005) 315 – 325, 2005

[13]    K. Keiser, R. Ramachandran, J. Rushing, H. Conover, S. Graves, "Distributed Services Technology for Earth Science Data Processing"

[14]    Landsat Imagery, http://www.landsat.org/ortho/index.php

[15]    M. Ehlers, M.A. Jadkowski, R.R. Howard, D.E. Brostuen, 1990, "Application of a remote sensing –GIS evaluation of urban expansion SPOT data for regional growth analysis and local planning", Photogrammetric Engineering & Remote Sensing, 56, 175 – 180, 1990

[16]    M. Hmida, Y. Slimani "Meta-learning in Grid-based Data Mining Systems", International Journal of Computer Networks & Communications (IJCNC) Vol.2, No.5, ISSN - [Online: 0974 - 9322] ISSN - [Print : 0975- 2293], 2010

[17]    M. Khemakhem, A. Belghith "Towards Trusted Volunteer Grid Environments", International Journal of Computer Networks & Communications (IJCNC) Vol.2, No.2, ISSN - [Online: 0974 - 9322] ISSN - [Print : 0975- 2293], 2010

[18]    M. Roth., T.R. Oke, W.J. Emery, "Satellite derived urban heat islands from three coastal cities and the utilization of such data in urban climatology", International Journal of Remote Sensing, 10, 1699 – 1720, 1989

[19] M. S Moran, T. R. Clarke, Y. Inoue, A. Vidal, "Estimating crop water deficit using the relation between surface-air temperature and precipitable water over the tropics, Journal and spectral vegetation index" Remote Sensing of Environment, 49, 246– 263, 1994

[20] P.M. Harris, S.J. Ventura, "The integration of geographic data with remotely sensed imagery to improve classification in an urban area", Photogrammetric Engineering & Remote Sensing, 61, 993 – 998, 1995

[21] P.M. Treitz, P.J. Howard, P. Gong, "Global change and terrestrial ecosystems: the operational plan. IGBP Report No.21", International Geosphere- Biosphere Programme, Stockhol, 1992.

[22] P.S. Chavez, "Image-based atmospheric correction—revisited and Improved", Photogrammetric Engineering and Remote Sensing, 62(9), 1025– 1036, 1996

[23] P.S. Chavez, "An improved dark-object subtraction technique for atmospheric scattering correction of multispectral data", Remote Sensing of Environment, 24:459-479, 1988

[24] Q. Li, H, Sui, Y. Feng, Q. Zhan, C. Xu, "Research On Remote Sensing Information Processing Services Based On Semantic Web Services", The International Archives of the Photogrammetry, Remote Sensing and Spatial Information Sciences. Vol. XXXVII, Part B4, Beijing 2008

[25] Q. Weng, D. Lu, J. Schubring, "Estimation of land surface temperature-vegetation abundance relationship for urban heat island studies", Remote Sensing of Environment 89 (4) 467–483, 2004

[26] Q. Weng, "Fractal analysis of satellite detected urban heat island effect", Photogrammetric Engineering & Remote Sensing, 69 (5), 555 – 566, 2003

[27] Remote Sensing Guide, http://cbc.rs-gis.amnh.org/

[28] T.N. Carson, R.R. Gillies, E.M. Perry, "A method to make use of thermal infrared temperature and NDVI measurements to infer surface soil water content and fractional vegetation cover", Remote Sensing of Environment, 9, 161 – 173, 1998

[29] Z. Wan, W. Snyder "MODIS land-surface temperature algorithm theoretical basis document (LST ATBD): version 3.2.,1996

## Authors

Cristina Serban is an Assistant Professor of the Civil Engineering at "Ovidius" University of Constanta. Drd. Serban obtained the degrees in 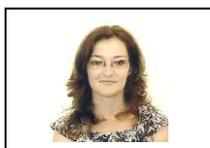 Mathematics and Computer Science at the Faculty of Mathematics and Computer Science ("Ovidius" University of Constanta). Drd. Serban is a Ph.D. student in Computer Science at the Faculty of Mathematics and Computer Science, West University of Timisoara. Drd. Serban's research field include Distributed Systems, Parallel Computing, Software Engineering. She published more than 20 articles and was member in 4 scientific research projects.

Carmen Elena Maftei is a Professor of the Civil Engineering Faculty at "Ovidius" Universty of 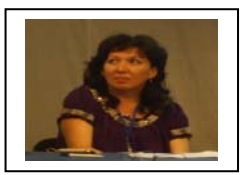 Constanta and Head of the Research Department of the same university. Born and raised in Romania, Dr. Maftei was educated at Arts High School in Buzau. She obtained the degrees in Civil Engineering at Polytechnic Institute in Iasi (Technical University "Ghe. Asachi" today). She received a Ph.D. in Civil Engineering in 2002, both from Montpellier Academy (University II) and "Ovidius" University of Constanta. Dr. Maftei's main research focus is on Hydrology and Hydrogeology. She is currently an associate editor of International Journal of Ecology and Development and Romanian Civil Engineering Journal. Since 1993 Dr. Maftei published 7 books in her field research and more than 60 articles. She conducted 7 research projects and she was member in other 11 projects. Dr. Maftei is member of several professional societies (IAHS, WASWC, AIFCR and AGIR since 2003, IWRA and IAENG since 2007).